\documentclass[useAMS,usenatbib]{mn2e}
\usepackage{epsfig,natbib}
\usepackage{amssymb}
\usepackage{graphicx}
\usepackage{placeins}
\usepackage{times}
\usepackage{epsfig}

\newcommand\ion[2]{#1{\sc #2}}
\newcommand\fion[2]{$[$#1{\sc #2}$]$}

\title[ Galaxy counterparts of metal-rich DLAs II ] {Galaxy
  counterparts of metal-rich damped Ly$\alpha$ absorbers -- II. A
  solar-metallicity and dusty DLA at $z_{\rm
    abs}=2.58$\thanks{Based on observations carried out under prog.
    ID 084.A-0303 with the X-shooter spectrograph installed at the
    Cassegrain focus of the Very Large Telescope (VLT), Unit 2 --
    Kueyen, operated by the European Southern Observatory (ESO) on
    Cerro Paranal, Chile.}}
\author[J.~P.~U. Fynbo et al.]{J. P. U. Fynbo,$^{1}$\thanks{E-mail:
jfynbo@dark-cosmology.dk}
C. Ledoux,$^{2}$
P. Noterdaeme,$^{3}$
L. Christensen,$^{4}$
P. M\o ller,$^{5}$
A.~K. Durgapal,$^{6}$\newauthor
P. Goldoni,$^{7,8}$
L. Kaper,$^{9}$
J.-K. Krogager,$^{1}$
P. Laursen,$^{10}$
J.~R. Maund,$^{1,11}$
B. Milvang-Jensen,$^{1}$\newauthor
K. Okoshi,$^{12}$
P.~K. Rasmussen,$^{13}$
T.~J. Thorsen,$^{1}$
S. Toft$^{1}$ and
T. Zafar$^{1}$\\
$^{1}$Dark Cosmology Centre, Niels Bohr Institute, Copenhagen University, Juliane Maries Vej 30, 2100 Copenhagen O, Denmark\\
$^{2}$European Southern Observatory, Alonso de C\'ordova 3107, Vitacura, Casilla 19001, Santiago 19, Chile\\
$^{3}$Departamento de Astronom\'ia, Universidad de Chile, Casilla 36-D, Santiago, Chile\\
$^{4}$Excellence Cluster Universe, Technische Universit\"t M\"unchen, Bolzmanstrasse 2, 85748 Garching bei M\"unchen, Germany\\
$^{5}$European Southern Observatory, Karl-Schwarzschildstrasse 2, 85748 Garching bei M\"unchen, Germany\\
$^{6}$Department of Physics, DSB Campus Kumaun University, Nainital, Uttarakhand, India\\
$^{7}$Laboratoire Astroparticule et Cosmologie, 10 rue A. Domon et L. Duquet, 75205 Paris Cedex 13, France\\
$^{8}$Service d'Astrophysique, DSM/IRFU/SAp, CEA-Saclay, 91191 Gif-sur-Yvette, France\\
$^{9}$Astronomical Institute 'Anton Pannekoek', University of Amsterdam, Kruislaan 403, 1098 SJ Amsterdam, the Netherlands\\
$^{10}$Oskar Klein Centre, Department of Astronomy, Stockholm University, 10691 Stockholm, Sweden\\
$^{11}$Department of Astronomy and Astrophysics, University of California, Santa Cruz, CA 95064, USA\\
$^{12}$Tokyo University of Science, Oshamanbe, Hokkaido, 049-3514 Tokyo, Japan\\
$^{13}$Niels Bohr Institute, Copenhagen University, Juliane Maries Vej 30, 2100 Copenhagen O, Denmark
}

\begin{document}


\pagerange{1 -- 1} \pubyear{2010}

\maketitle


\begin{abstract}
This is the second paper in a series reporting on results from a
survey conducted with the ESO VLT/X-shooter spectrograph. We target
high metallicity damped Lyman-$\alpha$ absorbers (DLAs) with the aim of
investigating the relation between galaxies detected in emission and
those detected in absorption.
%
%
Here, we report on the discovery of the galaxy counterpart of the
$z_{\rm abs}=2.58$ DLA on the line-of-sight to the $z=3.07$ quasar
SDSS J\,091826.16$+$163609.0 (hereafter Q\,0918$+$1636). The galaxy
counterpart of the DLA is detected in the \fion{O}{iii} $\lambda
5007$ and \fion{O}{ii} $\lambda\lambda 3726,3729$ emission lines
redshifted into the NIR at an impact parameter of 2.0 arcsec (16 kpc
at $z=2.58$). Ly$\alpha$ emission is not detected down to a
$3\sigma$ detection limit of $5\times 10^{-18}$ erg s$^{-1}$
cm$^{-2}$, which, compared to the strength of the oxygen lines,
implies that Ly$\alpha$ emission from this galaxy is suppressed by
more than an order of magnitude. The DLA has one of the highest metallicities
measured so far at comparable redshifts. We find
evidence for substantial depletion of refractory elements onto dust
grains. Fitting the main metal line component of the DLA, which is
located at $z_{\rm abs}=2.5832$, 
we measure metal
abundances from \ion{Zn}{ii}, \ion{S}{ii}, \ion{Si}{ii},
\ion{Cr}{ii}, \ion{Mn}{ii}, \ion{Fe}{ii} and \ion{Ni}{ii} of
$-0.12\pm 0.05$, $-0.26\pm 0.05$, $-0.46\pm 0.05$, $-0.88\pm 0.05$,
$-0.92\pm 0.05$, $-1.03\pm 0.05$ and $-0.78\pm 0.05$, respectively.
In addition, we detect absorption in the Lyman and Werner bands of
molecular hydrogen (H$_2$), which represents the first detection of
H$_2$ molecules with X-shooter. 
The background quasar Q\,0918$+$1636 is amongst the
reddest QSOs at redshifts $3.02<z<3.12$ from the SDSS catalogue. Its
UV to NIR spectrum is well fitted by a composite QSO spectrum
reddened by SMC/LMC-like extinction curves at $z_{\rm abs}=2.58$
with a significant amount of extinction given by $A_V \approx 0.2$
mag. This supports previous claims that there may be more metal-rich
DLAs missing from current samples due to dust reddening of the
background QSOs. The fact that there is evidence for dust both in
the central emitting regions of the galaxy (as evidenced by the lack
of Ly$\alpha$ emission) and at an impact parameter of 16 kpc (as
probed by the DLA) suggests that dust is widespread in this system.
\end{abstract}

\clearpage


\section{Introduction}

The comparison between absorption-line selected and emission-selected
galaxies at redshifts $z>2$ has a long history \citep[e.g.,][and
references therein]{Smith89,Moller93,Wolfe05}. However, progress in
this field has been slow and, for many years, there has been little
overlap between observational samples
\citep[e.g.,][]{Fynbo99,Moller02,Colbert02,Kulkarni06}. Recently, some
progress has been made though in order to build the bridge between the
two populations. Emission-selected galaxies have been studied to much
deeper rest frame flux limits than a decade before
\citep[e.g.,][]{Sawicki06,Gronwall07,Rauch08,Ouchi08,Grove09,Reddy09,Cassata10}.
\citet[][hereafter F10]{Fynbo10} also presented the first results from
a small survey to search for the galaxy counterparts of metal-rich
damped Ly$\alpha$ absorbers \citep[DLAs; see][for a review]{Wolfe05}.
In F10, the detection of the galaxy counterpart of a high-metallicity DLA
at $z_{\rm abs}=2.354$ toward Q\,2222$-$0946 was presented. F10
described the strategy and sample selection of the survey in details.
Here, we reiterate that the candidates were selected amongst SDSS QSOs
based on the strengths of \ion{Si}{ii} and \ion{Fe}{ii} absorption
lines (i.e., not directly from Ly$\alpha$).


In this paper, we present new results based on observations of the
second target of the survey, the $z=3.07$ quasar SDSS
J\,091826.16$+$163609.0 (hereafter Q\,0918$+$1636). Q\,0918$+$1636 was
selected as its spectrum features a metal-rich absorber at $z=2.412$
having rest frame equivalent widths (EWs) of the targeted
\ion{Si}{ii}$\lambda 1526$ and \ion{Fe}{ii}$\lambda\lambda\lambda
2344,2374,2382$ lines of 2.1~\AA, 2.1~\AA, 1.4~\AA\ and 2.6~\AA,
respectively. The properties of this absorber will be the subject of
another paper describing a sample of objects. For now, we note that
this absorber, which is also a DLA, has a high metallicity with
[Si/H$]=-0.6$. Here, we describe the serendipitous discovery and
properties of a second DLA, at $z_{\rm abs}=2.5832$, along the
line-of-sight to Q\,0918$+$1636. This system happens to also fulfill
our sample selection criteria, with rest frame EWs of 2.4~\AA,
2.4~\AA, 1.6~\AA\ and 3.0~\AA\ for \ion{Si}{ii}$\lambda 1526$ and
\ion{Fe}{ii}$\lambda\lambda\lambda 2344,2374,2382$, respectively. The
\ion{Fe}{ii} lines from this system are redshifted outside of the
wavelength interval suited for automatic searches of absorption lines
in SDSS spectra and hence were not identified. Interestingly also,
neither of the two DLAs in this QSO spectrum have been identified in
automatic DLA (i.e., Ly$\alpha$-based) searches in SDSS
\citep{Prochaska05,Noterdaeme09a}. 
This is due to a recently identified bias against the detection of damped
Lyman-alpha absorption lines at the blue end of the spectra
\citep[see][]{Noterdaeme09a}.  Indeed, the signal-to-noise ratio is decreased
significantly by the presence of any strong and/or several closeby DLA
lines.

Throughout this paper, we assume a flat cosmology with
$\Omega_{\Lambda}=0.70$, $\Omega_m=0.30$ and a Hubble constant of
$H_0=70$~km~s$^{-1}$~Mpc$^{-1}$.

\section{Observations and data reduction}

Q\,0918$+$1636 was observed on February 16 2010 with X-shooter at the
VLT. The QSO was observed at three position angles (PAs), namely
$60^\mathrm{o}$, $-60^\mathrm{o}$ and $0^\mathrm{o}$ (all East of
North). The purpose of using three slit positions is to cover a larger
field of view around the QSO as shown in fig.~1 of F10. The
integration time at each PA was 3600 s, and a 1.3$\arcsec$-wide slit
in the UVB arm and 1.2$\arcsec$-wide slits in the VIS and NIR arms
were used.

The expected resolving power with the above setup is 4000, 6700 and
4300 in the UVB, VIS and NIR arms, respectively. This is confirmed by
the width of sky emission lines in the spectra. However, the seeing
during the observations was significantly smaller than the widths of
the slits (i.e., 0.67 arcsec in all three exposures as measured from
the width of the QSO trace around 7500~\AA) and hence the true
spectral resolution is higher than that measured from sky emission
lines. In the VIS arm, we measure the resolution directly from the
width of telluric absorption lines. We then assume that the ratio
between expected and true resolutions is the same in all three
spectroscopic arms, and the degradation of seeing as a function of
decreasing wavelength is calculated as the ratio of the wavelengths to
the power of 0.2. In this way, we infer resolving power values of
6400, 11900 and 8800 in the UVB, VIS and NIR arm, respectively.

We processed the spectra using the X-shooter data reduction pipeline
\citep[see][see also F10]{Goldoni06}. First, raw frames were corrected
for the bias level (UVB and VIS) or dark current (NIR). Then, after
background subtraction, cosmic ray hits were detected and removed
using the method developed by \citet{Vandokkum01} while sky emission
lines were subtracted using the method described in \citet{Kelson03}.
After division by a master flat-field, the spectral orders were
extracted and rectified in wavelength space using a wavelength
solution previously obtained from calibration frames. The orders were
then merged and in their overlapping parts the merging was weighted by
the corresponding errors which were propagated in the process. From
the resulting merged 2D spectrum, a one-dimensional spectrum of the
QSO was extracted. This 1D spectrum together with its error file and
bad-pixel map are the final products of the reduction. Intermediate
products such as the sky spectrum and individual echelle orders (with
errors and bad-pixel maps) were also produced. The spectra were
flux-calibrated using a spectrophotometric standard star observed
during the same night. The flux calibration was checked against the
flux-calibrated SDSS QSO spectrum and found to be consistent with it.

\section{Results}

\subsection{Emission properties of the DLA-galaxy counterpart}

\begin{figure}
\includegraphics[width=0.48\textwidth]{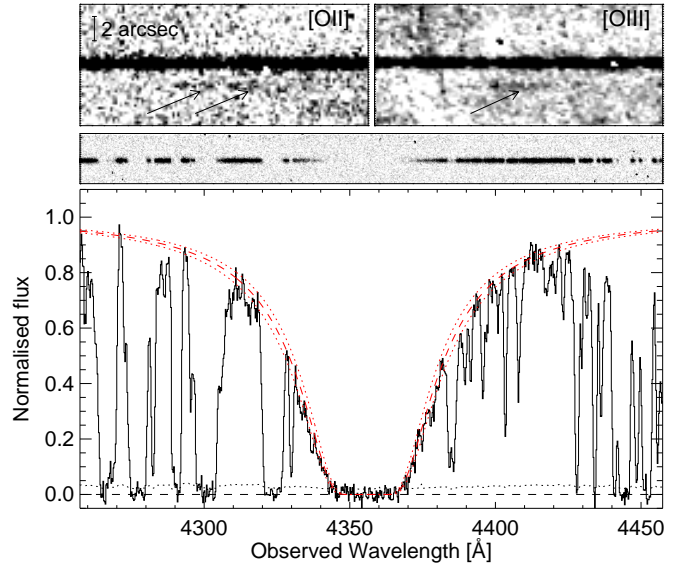}
\caption{{\it Top panels:} The detected \fion{O}{ii} $\lambda\lambda
  3726,3729$ and \fion{O}{iii} $\lambda 5007$ emission lines from the
  DLA-galaxy counterpart observed 2.0 arcsec away from the QSO
  continuum. The arrows point to the predicted centroids of the lines
  if these were at the same redshift as the low-ionisation absorption
  lines from the DLA. {\it Middle and bottom panels:} The region
  around Ly$\alpha$ from both the two- and one-dimensional spectra
  with PA$=60^\circ$. As can be seen, the Ly$\alpha$ emission line is
  not detected. The red dashed and dotted lines show the Voigt profile
  fit to the damped Ly$\alpha$ line and associated $1\sigma$
  uncertainties, respectively, corresponding to a neutral hydrogen
  column density of $\log N($\ion{H}{i}$)=20.96\pm
  0.05$.\label{fig:dla2d}}
\end{figure}

We do not detect Ly$\alpha$ emission at any of the three position
angles (see Fig.~\ref{fig:dla2d}). We calculate a conservative
3$\sigma$ detection limit from a 1000~km\,s$^{-1}$ $\times$
2.0 arcsec square aperture inside the trough of the $z=2.58$ damped
Ly$\alpha$ absorption line. The velocity width of 1000~km\,s$^{-1}$
is based on the Ly$\alpha$ line from the galaxy counterpart of the
DLA towards Q\,2222$-$0946 (F10, see also theoretical profiles in
\citet{Laursen10}), whereas the spatial width of 2 arcsec is based 
on the fact that Ly$\alpha$ emission can be quite extended 
\citep[e.g.,][]{Moller98,Fynbo03,Rauch08}. 
The resulting 
limit is $5\times 10^{-18}$ erg s$^{-1}$ cm$^{-2}$.
Assuming that all Ly$\alpha$ photons
escape, case-B recombination and the relation between H$\alpha$
luminosity and star formation rate (SFR) from \citet{Kennicutt98}, we
would place an upper limit on the SFR of 0.3 M$_{\sun}$ yr$^{-1}$.
Unfortunately, the expected position of the H$\alpha$ emission line is
in a part of the NIR spectrum where the sky background is too high to
allow for a useful detection limit to be derived. However, we do
detect the \fion{O}{iii} $\lambda 5007$ emission line in the
PA$=60^\circ$ spectrum at an impact parameter of 2.0 arcsec (see
Fig.~\ref{fig:dla2d}). The redshift of the \fion{O}{iii} $\lambda
5007$ line is consistent with that of low-ionisation metal lines to
within an uncertainty of about 50~km\,s$^{-1}$. This line is detected
neither in the PA$=-60^\circ$ nor the PA$=0^\circ$ spectra. We also
detect both components of the \fion{O}{ii} $\lambda\lambda 3726,3729$
doublet but with a lower signal-to-noise ratio and in a region
somewhat affected by telluric absorption lines. The flux of the
\fion{O}{iii} $\lambda 5007$ line is $f=1.7\pm 0.2\times 10^{-17}$ erg
s$^{-1}$ cm$^{-2}$, which corresponds to a luminosity of $L=9\times
10^{41}$ erg s$^{-1}$. This is a lower limit due to the possibility of
slit-loss. For the \fion{O}{ii} doublet, we infer a total flux of
about $2.5\times 10^{-17}$ erg s$^{-1}$ cm$^{-2}$ after correcting for
telluric absorption. Using the relation from \citet{Kennicutt98}, we
infer a SFR of about 20 M$_{\sun}$ yr$^{-1}$. Hence, Ly$\alpha$
emission from this system appears to be suppressed by more than an
order of magnitude.

In order to exclude the possibility of the existence of a galaxy
counterpart located at small impact parameter, we also performed
spectral point spread function subtraction as in F10. No emission line
is detected at an impact parameter smaller than 2 arcsec. Any line
with a flux similar to or larger than the detected \fion{O}{iii} line
would be detected in the data.

\subsection{Absorption-line properties of the DLA}

\subsubsection{Metal lines}

The total \ion{H}{i} column density of the system is well constrained
by the damped Ly$\alpha$ absorption line to be $\log
N($\ion{H}{i}$)=20.96\pm 0.05$ (see Fig.~\ref{fig:dla2d}), where the
error is a conservative error on the uncertainty, which is dominated
by the systematic error from the normalisation. In particular, the
fact that the DLA line is located in the red wing of the \ion{O}{vi}
emission line of the QSO complicates the normalisation somewhat.

\begin{figure}
\psfig{file=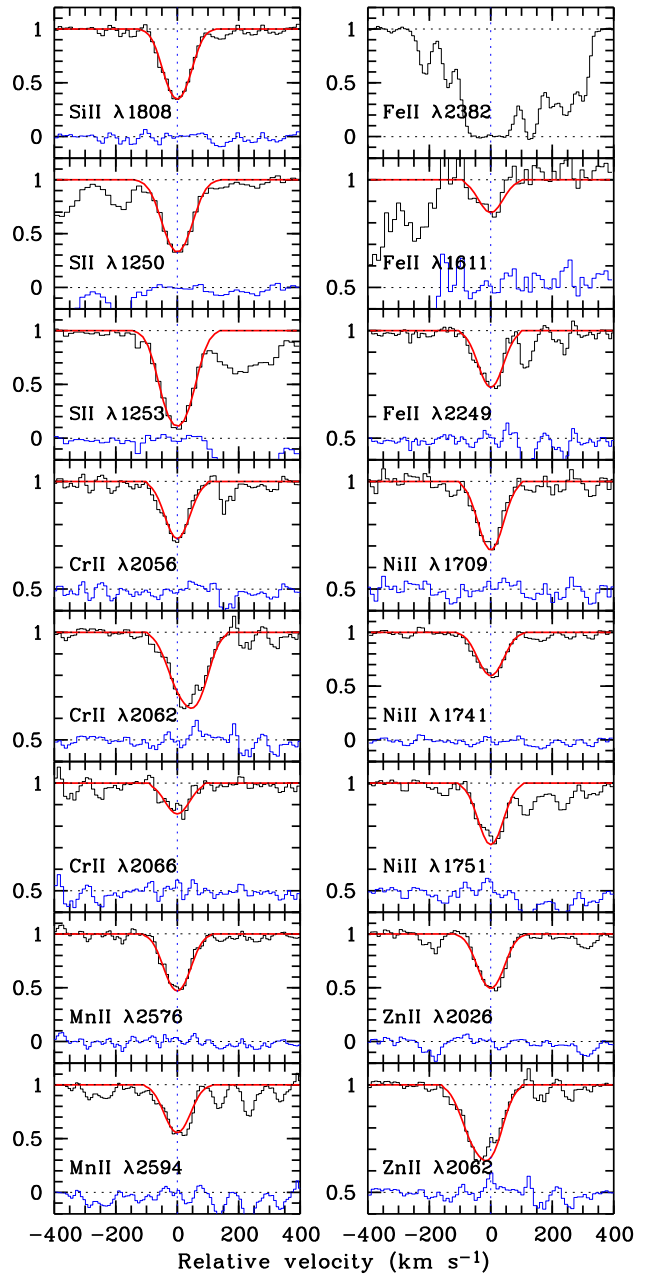,scale=0.69,clip=}
\caption{Results of Voigt-profile fits to low-ionisation lines from
  the $z_{\rm abs}=2.5832$ DLA toward Q\,0918$+$1636. Most of the
  absorption is concentrated within one velocity component. In order
  to derive overall metallicities in the neutral gas phase, we focus
  on fitting this main component which is responsible for at least
  85\% of the total metal column density (see text). The Voigt-profile
  fits are shown with a red line and the residuals from the fits are
  shown with a blue line (for display purposes, a constant of 0.5 has
  been added to the residuals in some of the panels).
  \label{fig:metallines} }
\end{figure}

The absorption profiles of the strongest \ion{Fe}{ii} and \ion{Mg}{ii}
lines span a velocity range of up to 600~km\,s$^{-1}$ (see
\ion{Fe}{ii}$\lambda 2382$ in the upper right panel of
Fig.~\ref{fig:metallines}). However, most of the absorption as seen
from unsaturated or weakly saturated lines from singly ionized species
is concentrated within a single broad velocity component at $z_{\rm
  abs}=2.5832(2)$. In order to derive overall metallicities in the
neutral gas phase, we focus on fitting this main component. For the
fitting, we use the package FITLYMAN as available in MIDAS. From the
weak \ion{Si}{ii}$\lambda 1808$ line, one can see that this main
component is responsible for 85\% of the total EW and hence most
likely -- at least -- 85\% of the total column density of
low-ionisation species.

The results of Voigt-profile fitting are summarised in
Table~\ref{metallines} and displayed in Fig.~\ref{fig:metallines}.
The instrumental resolution FWHM is 47 and 25~km\,s$^{-1}$ for the
UVB- and VIS-arm spectra, respectively. With a broadening parameter
$b=52$~km\,s$^{-1}$, the fitted metal-line profile is well-resolved.
However, should it include in reality one or several narrow
components, hidden saturation could be an issue and strictly speaking
our measurements should be considered as lower limits. This means that
the overall metallicity of this DLA could be even higher than solar.
However, apart from \ion{S}{ii} and to some extent also \ion{Si}{ii},
we are considering relatively weak absorption lines in the fitting
process (see Fig.~\ref{fig:metallines}) so that the measured column
densities for the corresponding species should depend weakly on
individual $b$ values if at all.

The usual complications of line blending in low-resolution spectra are
avoided here as a single well-defined component is fit and therefore
the presence of blending with unrelated absorption line features is
easier to identify. Absorption from neutral carbon is detected
(\ion{C}{i}$\lambda\lambda 1560,1656$) but due to blending with
\ion{C}{i}$^\star$ and \ion{C}{i}$^{\star\star}$ lines neutral carbon
is not considered in the analysis any further. Also, it is not
  possible to infer whether \ion{C}{ii}$^{\star}$ is present due to
  blending with the \ion{C}{ii}$\lambda 1334$ line which is highly
  saturated. Hence, we cannot determine the relative strengths of
  \ion{C}{ii}$^{\star}$ and \ion{C}{ii} and use them as diagnostics to
  infer the presence of local radiation fields \citep{Wolfe08}. The
errors given in Table~\ref{metallines} are the formal statistical
errors from FITLYMAN. The systematic errors from, e.g., normalisation
and not knowing the instrumental resolution precisely, are larger than
this. An error of 0.05 dex would be a conservative estimate including
all error sources.

The metallicity of the system is exceptionnally high. From
\ion{Zn}{ii}, which is little depleted onto dust grains
\citep{Meyer90}, we infer a metallicity of $-0.12\pm 0.05$. Note that
the splitting between \ion{Zn}{ii}$\lambda 2026$ and
\ion{Mg}{i}$\lambda 2026$ is 50~km\,s$^{-1}$ implying that \ion{Mg}{i}
cannot significantly contribute to the \ion{Zn}{ii} column density.
Here, we adopt the solar photosphere abundances from
\citet{Asplund09}. Given that we only fitted the main velocity
component, the overall metallicity of the DLA must be solar or higher
than solar in case of hidden saturation. We also note that the [Mn/Fe]
ratio is larger than solar, which is only seen at or above solar
metallicity \citep{Ledoux02}. For elements which are sensitive to dust
depletion \citep[e.g.,][]{Meyer90,Pettini97,Ledoux02}, we find
[Fe/H$]=-1.03$, [Ni/H$]=-0.78$, [Mn/H$]=-0.92$, and [Cr/H$]=-0.88$
implying substantial dust depletion for these elements.

\begin{figure}
\includegraphics[width=0.48\textwidth]{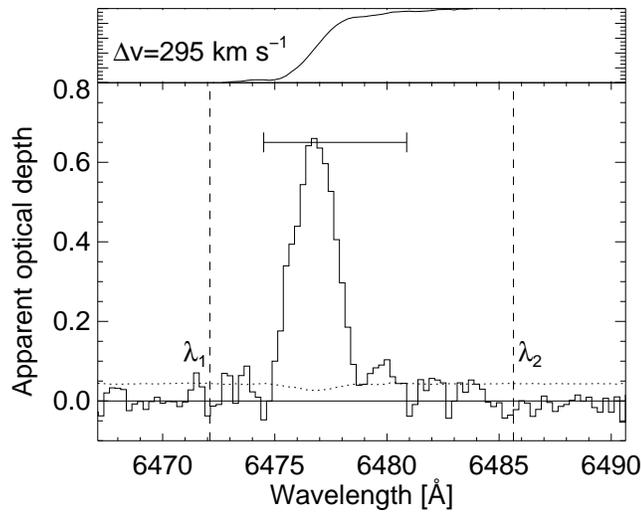}
\caption{The profile of the \ion{Si}{ii} $\lambda 1808$ line. The
  profile consists of a main, broad component and a wing towards the
  red end of the profile. The velocity width of the line, measured
  following the method of Ledoux et al. (2006), is 295~km\,s$^{-1}$
  (marked with a horizontal segment). $\lambda_1$ and
  $\lambda_2$ are the start and end wavelengths used to integrate the
  profile. \label{fig:profile}}
\end{figure}

To compare the kinematics of the absorption-line profiles with that of
other DLAs, we follow the procedure of \citet{Ledoux06} and calculate
the line-profile velocity width, $\Delta V$, as
$c\left[\lambda(95\%)-\lambda(5\%)\right]/\lambda_0$, where
$\lambda(5\%)$ and $\lambda(95\%)$ are the wavelengths corresponding
to, respectively, the 5 and 95 percentiles of the apparent optical
depth distribution, and $\lambda_0$ is the first moment (the average)
of this distribution \citep[see fig.~1 of][]{Ledoux06}. We again
choose the \ion{Si}{ii} $\lambda$1808 transition as it is a
low-ionisation transition and the line is only mildly saturated. The
apparent line optical depth and the derived velocity width are shown
in Fig.~\ref{fig:profile}. We infer a velocity width of 295 km
s$^{-1}$ in good agreement with the velocity-metallicity relation for
DLAs \citep{Ledoux06}.

\subsubsection{Molecular absorption}

\begin{figure}
\centering
\begin{tabular}{c}
\includegraphics[width=\hsize]{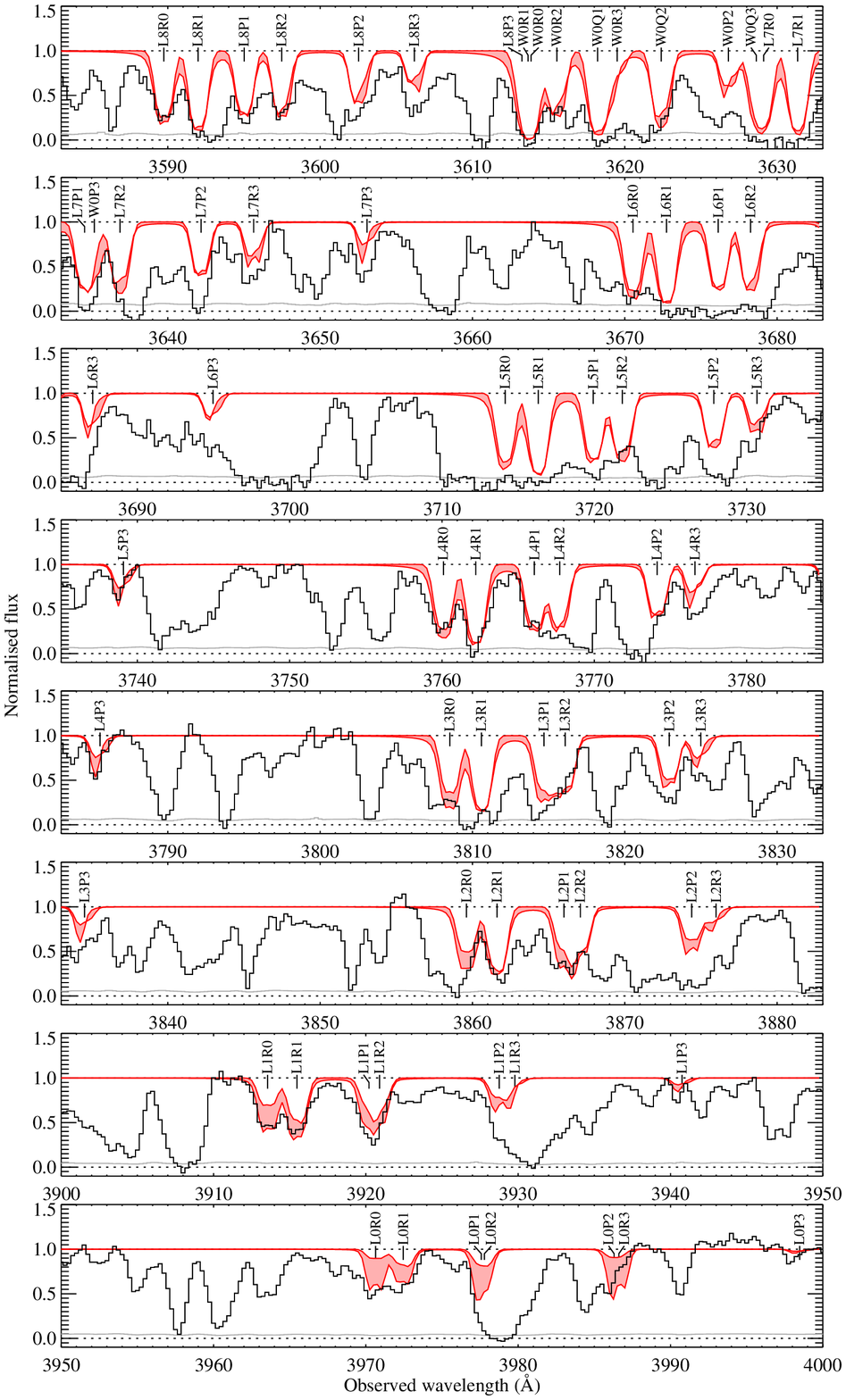}\\
\includegraphics[width=\hsize]{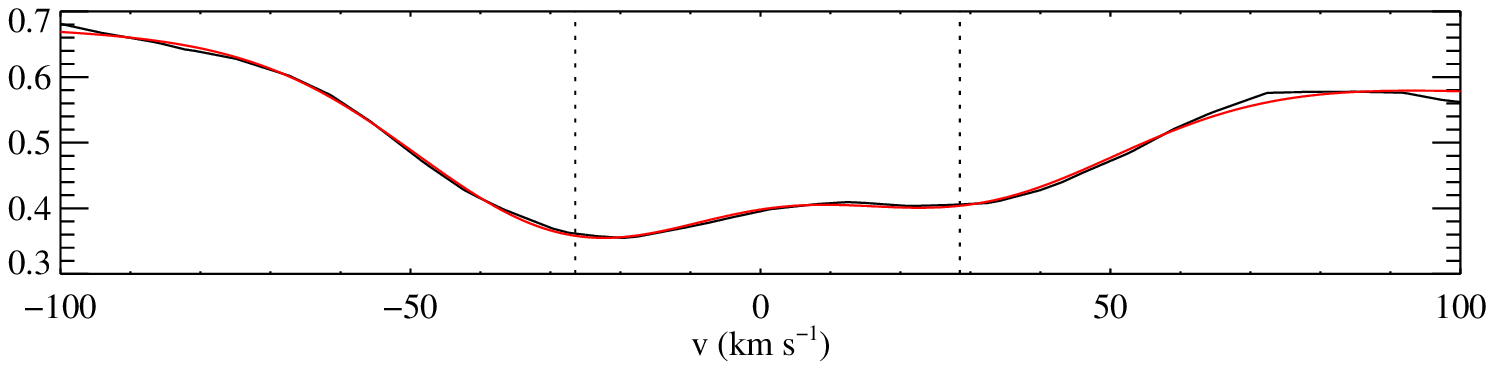}\\
\end{tabular}
\caption{\label{figH2} Portion of the X-shooter spectrum featuring
  H$_2$ absorption lines superimposed on the Ly$\alpha$ forest. The
  label over each absorption line indicates the band (L: Lyman, W:
  Werner), the vibrational level of the upper state, the branch (P, Q
  or R for $\Delta J=+1,0,-1$, respectively) and the rotational level
  J of the lower state. The synthetic profiles corresponding to $b=1$
  and $b=20$~km\,s$^{-1}$ are overplotted as a shaded area. The
    bottom panel represents the stacking of unsaturated, unblended
    lines (L0R0, L1R0, L1R1, L4R0 and L8R0), revealing better the two
    peaks in the H2 velocity profile.}
\end{figure}

Absorption lines from the Lyman and Werner bands of molecular hydrogen
(H$_2$) are detected at $z=2.5832$ in the X-shooter spectrum of
Q\,0918$+$1636 (see Fig.~\ref{figH2}). The observed velocity extent of
the H$_2$ profile is wider than the instrumental resolution FWHM (48
km s$^{-1}$ at 3800 \AA) and much wider than the typical Doppler
parameter of individual molecular lines as measured from
high-resolution spectroscopy \citep[usually
$b\sim3$~km\,s$^{-1}$;][]{Ledoux03}. This indicates that the actual
H$_2$ profile comprises several blended components. Two absorption
peaks separated by about 55~km\,s$^{-1}$ can be distinguished in the
H$_2$ profile, while they are not seen in the profile of metal lines
that are fitted with a single component having $b=52.1$~km\,s$^{-1}$.
This means that the metals are somehow continuously spread over
FWHM~$=b\times 2\sqrt({{\rm ln} \,2})\sim 87$~km\,s$^{-1}$, while the
H$_2$ components are discrete and well-separated in velocity space.
Indeed, this is a general behaviour as seen in other H2-bearing
systems studied at high spectral resolution
\citep[e.g.][]{Ledoux02b}


In order to estimate the H$_2$ column density, we compared the data
with synthetic profiles of H$_2$ absorption consisting of two
components separated by 55~km\,s$^{-1}$ built with the same spectral
resolution and binning. We varied the column density in rotational
levels from $J=0$ to $J=3$ for each component until we got a synthetic
spectrum consistent with the data. We repeated this exercise for a
wide range of Doppler parameters ($b=~1$ to 20~km\,s$^{-1}$) to get a
realistic range on the total H$_2$ column density:
$N($H$_2$)=1.5$\times$10$^{16}$ -- 1.1$\times$10$^{19}$\,cm$^{-2}$.
This corresponds to an overall molecular fraction $f=
(2N($H$_2)/(2N($H$_2)$+$N($H\,{\sc
  i}$)))=3.3\times10^{-5}$--$2.4\times10^{-2}$.

Details on the column densities in each rotational level and for the
two components are given in Table~\ref{tabH2}. These measurements are
only indicative and should be considered with caution because of the
various uncertainties. Only high spectral resolution observations are
suitable to properly resolve the H$_2$ line profile in individual
components, measure accurately the column densities, and study the
physical conditions in the gas
\citep[e.g.,][]{Srianand05,Noterdaeme07}. Nevertheless, the presence
of H$_2$ in this system is firmly established.

\begin{table}
  \caption {Ionic column densities in the main metal-line component of
    the $z_{\rm abs}=2.5832$ DLA toward Q\,0918$+$1636.\label{metallines}}
\begin{center}
\begin{tabular}{lllll}
\hline
\hline
Ion & Transition & $\log N\pm\sigma _{\log N}$ &  [M/H]  & $b\pm\sigma _b$\\
    & lines used &                             &         & (km\,s$^{-1}$) \\
\hline
\ion{Si}{ii} & 1808             & 16.01$\pm$0.01  &  $-$0.46$\pm$0.05 & 52.1$\pm$0.5  \\
\ion{S}{ii}  & 1250,1253        & 15.82$\pm$0.01  &  $-$0.26$\pm$0.05 & 52.1$\pm$0.5  \\
\ion{Cr}{ii} & 2056,2062,2066   & 13.72$\pm$0.01  &  $-$0.88$\pm$0.05 & 52.1$\pm$0.5  \\
\ion{Mn}{ii} & 2576,2594        & 13.47$\pm$0.01  &  $-$0.92$\pm$0.05 & 52.1$\pm$0.5  \\
\ion{Fe}{ii} & 1611,2249        & 15.43$\pm$0.01  &  $-$1.03$\pm$0.05 & 52.1$\pm$0.5  \\
\ion{Ni}{ii} & 1709,1741,1751   & 14.40$\pm$0.01  &  $-$0.78$\pm$0.05 & 52.1$\pm$0.5  \\
\ion{Zn}{ii} & 2026,2062        & 13.40$\pm$0.01  &  $-$0.12$\pm$0.05 & 52.1$\pm$0.5  \\
\hline
\end{tabular}
\end{center}
\end{table}



\begin{table}
\centering
\caption{\label{tabH2} H$_2$ column densities. The velocities, $\Delta v$,
  of the two components are given in km\,s$^{-1}$ with respect to
  $z_{\rm abs}=2.5832$.}
\begin{tabular}{c c c}
\hline
\hline
Rot. level & \multicolumn{2}{c}{$\log N$~(cm$^{-2}$)} \\
 component & $b$=1~km\,s$^{-1}$ & $b$=20~km\,s$^{-1}$ \\
\hline
$J=0$     &   18.45                &  15.4                 \\
$\Delta\,v=-26$    &   18.2                 &   15.1                 \\
$\Delta\,v=+28$    &   18.1                 &   15.1                 \\
\hline
$J=1$     &   18.75                &  15.85                \\
$\Delta\,v=-26$    &   18.5                 &   15.5                 \\
$\Delta\,v=+28$    &   18.4                 &   15.6                 \\
\hline
$J=2$     &   18.35                & 15.6                  \\
$\Delta\,v=-26$    &   18.2                 &  15.3                  \\
$\Delta\,v=+28$    &   17.8                 &  15.3                  \\
\hline
$J=3$     &   17.7                 &   15.0                 \\
$\Delta\,v=-26$    &   17.7                 &  14.8                  \\
$\Delta\,v=+28$    &   16.4                 &  14.6                  \\
\hline
Total   &   19.05                &  16.15                 \\
\hline
\end{tabular}
\end{table}

\section{Discussion and conclusions}

\subsection{Nature of the DLA galaxy counterpart}

The $z_{\rm abs}=2.5832$ DLA toward Q\,0918$+$1636 is amongst the
highest metallicity DLAs known to date \citep[compared, e.g., to the
sample of][]{Kaplan10}. The fact that we detect the galaxy counterpart
of this DLA is consistent with our suggestion that high metallicity
DLAs should have bright galaxy counterparts
\citep{Moller04,Fynbo08,Fynbo10}. Indeed, stellar mass, metallicity
and star formation rate in galaxies are intimately related
\citep{Mannucci10}. Galaxy mass-metallicity relations have been found
to exist up to at least $z=3.5$
\citep[e.g.,][]{Tremonti04,Savaglio05,Erb06,Kewley08,Maiolino08,Lamareille09}.
The mass-metallicity relations at high redshift are steeper and offset
towards higher stellar masses for a given metallicity compared to the
relation observed in the local Universe \citep[see, e.g., fig.~8
in][]{Maiolino08}. Therefore, by selecting high-metallicity systems we
should pick the most massive structures. In turn, because the star
formation rate in galaxies is at any redshift correlated with total
stellar mass \citep[see fig.~1 in][and references therein]{Dutton10},
the most massive galaxies have expected star formation rates high
enough (SFR$\ga 10$ M$_\odot$ yr$^{-1}$) to be detected with our
observing method. It is encouraging that we have already found a
second case after F10 in agreement with this reasoning.

The large impact parameter of 2.0 arcsec, corresponding to 16.0 kpc,
is also consistent with the expectations
from a simple model \citep[see fig.~3 in][]{Fynbo08}. The detection of
H$_2$ absorption is consistent with the previously found trend that
H$_2$ is predominantly found in high-metallicity \citep{Petitjean06}
and dusty \citep{Ledoux03,Noterdaeme08} systems. Unlike the system
studied in F10, Ly$\alpha$ emission from the $z_{\rm abs}=2.5832$ DLA
galaxy toward Q\,0918$+$1636 is strongly suppressed. A plausible
reason for this is dust obscuration given the high metallicity and the
evidence for dust in the galaxy along the line-of-sight to the
background QSO \citep{Charlot91}. Whereas we have no direct
information on the amount of dust in the centre of the galaxy,
presumably traced by the centroid of the line emission, it is likely
to be higher than in the outer parts of the galaxy traced by the DLA.

\subsection{Ly$\alpha$ escape fraction}

Whereas the limit we infer for the Ly$\alpha$ luminosity is a similar to the
mean Ly$\alpha$ luminosity for DLA galaxies inferred by \citet{Rahmani10}
\citep[see also][]{Rauch10} it is still very low compared to the strength of
the [\ion{O}{iii}] and [\ion{O}{ii}] emission lines from the galaxy.

The escape fraction $f_{\mathrm{esc}}$ of Ly$\alpha$ photons at high
redshift is a highly coveted quantity as a plethora of interesting
physical properties of galaxies can be revealed by studying this line.
A number of authors have attempted to assess this matter, yielding
quite different results; the standard procedure has been to compare
the ratio of Ly$\alpha$-inferred SFRs to those of other proxies with
what is expected theoretically assuming case-B recombination, e.g.,
\citet[][a universal 2\% from comparing observational data with
modelled galaxies]{Ledelliou05,Ledelliou06}, \citet[][30\% from
comparing simulated galaxies with observed luminosity
functions]{Dayal09}, and \citet[][80\% from comparing
Ly$\alpha$-inferred SFRs to spectral energy distribution (SED)
modelling of observed Ly$\alpha$ emitters]{Gawiser06}.
\citet{Laursen09b}, using the Ly$\alpha$ radiative transfer code {\sc
  MoCaLaTA} \citep{Laursen09a}, calculated the full radiative transfer
equations in simulated galaxies, obtaining an anticorrelation between
$f_{\mathrm{esc}}$ and galactic size, from order unity to order a few
per cent.

Similarly, although \citet{Hayes10} found that an \emph{average}
$f_{\mathrm{esc}} = 5.3$\% adequate to explain their observations,
their fig.~3 shows a trend of $f_{\mathrm{esc}}$ decreasing with
$E(B-V)$. Comparing the inferred SFR from \fion{O}{ii} with that of
Ly$\alpha$, an upper limit of $\sim 1$\% is inferred for the galaxy
counterpart of the DLA studied here. The extinction is found to be
$A_V \simeq 0.2$ (see below), corresponding to a colour excess of
$E(B-V) \simeq 0.06$. The extinction in the centre of the galaxy is
likely to be higher. Furthermore, in calculating $E(B-V)$
\citet{Hayes10} assume a metallicity of 1/3 solar, whereas the present
galactic system was shown to have at least solar metallicity, thus
probably resulting in higher extinction. Another possible cause could
be that SFRs inferred from \fion{O}{ii} are after all associated with
rather large uncertainties, as the mean \fion{O}{ii}/H$\alpha$ ratios
vary substantially from galaxy to galaxy, up to an order of magnitude
\citep{Gallagher89,Kennicutt92}.




\subsection{Dust reddening}

\begin{figure*}
\includegraphics[width=1.01\textwidth]{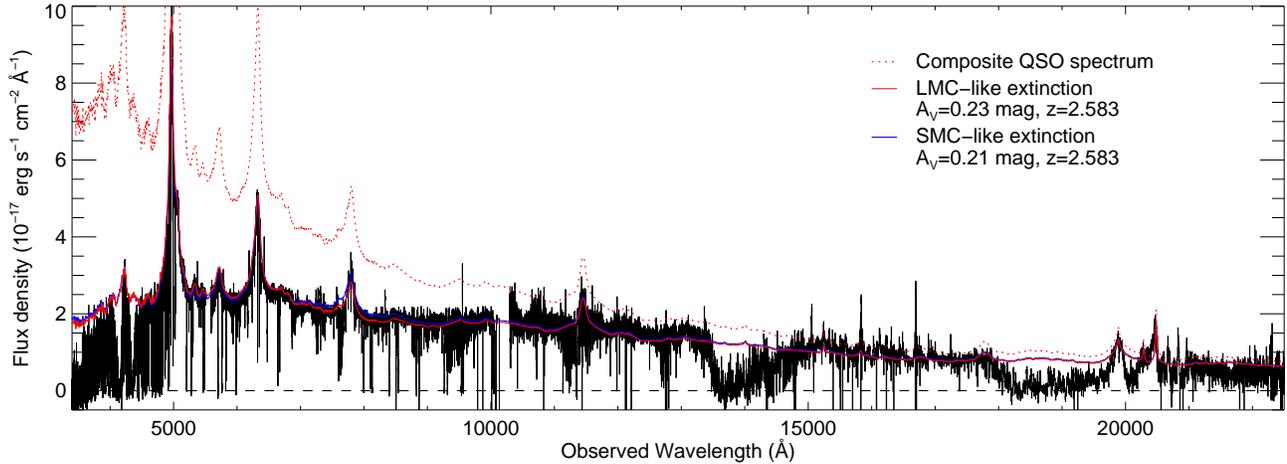}
\caption{The spectrum of Q\,0918$+$1636 after flux calibration. The
overall shape of the spectrum is well fitted by the composite QSO
spectrum from \citet{Telfer02}. In the figure, the unreddened
composite spectrum is shown with a dashed line, and the same
spectrum reddened by SMC- and LMC-like extinction curves with rest
frame $A_V=0.2$ mag is shown with solid red and blue lines,
respectively. The inferred dust-to-gas ratio is between those of the
LMC and the MW. The spectrum has been corrected for galactic
extinction with $E(B-V)=0.025$ from
\citet{Schlegel98}.\label{fig:extincA}}
\end{figure*}

\begin{figure}
\includegraphics[width=0.45\textwidth]{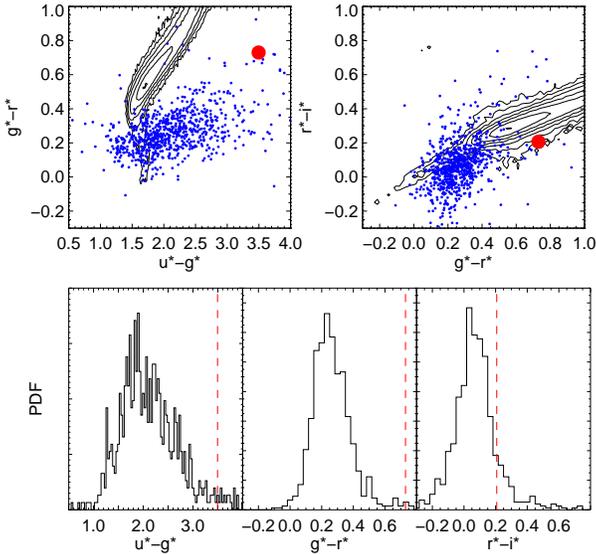}
\caption{Here, we compare the colours of Q\,0918$+$1636 with the
colours of other SDSS QSOs within $\Delta z=0.05$ of the redshift of
Q\,0918$+$1636 (BAL QSOs excluded). The top two panels show
colour-colour diagrams and the lower three panels show the
histograms of colours. The contours in the upper panels show the
colour distribution of the stellar locus. Q\,0918$+$1636 (marked as
a circle in the colour-colour plots and a dashed line in the
histograms) is among the reddest of all SDSS QSOs in this redshift
range (4\% in $u^*$$-g^*$, 0.7\% in $g^*$$-r^*$ and 12\% in
$r^*$$-i^*$).
\label{fig:extincB} }
\end{figure}

In Fig.~\ref{fig:extincA}, we show the X-shooter spectrum of
Q\,0918$+$1636. With a dashed line we show the composite QSO spectrum
from \citet{Telfer02} and it is evident that Q\,0918$+$1636 is redder
than the composite spectrum. The evidence for dust in the system comes from the observed
strong depletion of refractory elements. In addition, the spectrum of
Q\,0918$+$1636 appears substantially redder than the spectra of
typical QSOs around $z=3$. Q\,0918$+$1636 was selected as a candidate
high-$z$ QSO based on its red $u^*$$-g^*$ colour. It falls outside the
selection criteria for candidate $z<3.6$ QSOs in SDSS
\citep{Richards03}. In Fig.~\ref{fig:extincB}, we compare the colours
of Q\,0918$+$1636 with the colours of QSOs with redshift within
$3.07\pm 0.05$ from the SDSS DR7 QSO catalogue after removing BAL QSOs
\citep{Schneider07}. As seen, the distributions of QSO colours are
relatively narrow. However, the distributions have small extensions
towards redder colours and Q\,0918$+$1636 is among the reddest 4\%,
0.7\% and 12\% of QSOs at these redshifts for $u^*$$-g^*$, $g^*$$-r^*$
and $r^*$$-i^*$, respectively. With contours we show the colours of
the stellar locus from stripe 82 in the same colours
\citep{Ivezic07}. Only in the $u^*$$-g^*$ colour has the reddening
moved Q\,0918$+$1636 further away from the stellar locus and in these
bands the QSO is very faint. Hence, it is very difficult to select
reddened QSOs at these redshifts using optical colours. In a future
study, we will address how one can select even redder QSOs using
photometry in the near-IR bands. The most likely cause of the red
colour of Q\,0918$+$1636 is dust in the $z_{\rm abs}=2.5832$ DLA given
its high metallicity and strong evidence for depletion of refractory
elements on dust grains. Dust in the $z=2.412$ DLA could also
contribute, but as mentioned in the introduction this DLA not as metal
rich as the $z_{\rm abs}=2.5832$ DLA. The presence of dust along the
line-of-sight is supported by the fact that the SED of Q\,0918$+$1636
can be well fitted with the composite QSO spectrum from
\citet{Telfer02} reddened by SMC/LMC-like extinction at $z_{\rm
  abs}=2.5832$ with a total amount of extinction given by $A_V = 0.2$
mag. Here, we have used the prescription given in \citet{Pei92} to
model the extinction curve. The agreement with the observed spectrum
is very good. It is not possible to get a good fit with 
the MW extinction curve. 

The corresponding extinction-to-gas ratio, $A_V/N($H\,{\sc i})~$\approx
2.5\times10^{-22}$~mag\,cm$^2$, is in between the ratios inferred from LMC and
MW sightlines \citep[e.g.][]{Gordon03} and substantially higher than what is
found on average for DLAs \citep[$A_v/N($H\,{\sc i})~$\sim$~2-4$\times
10^{-23}$~mag\,cm$^2$][]{Vladilo08}.
The fact that reddening has moved Q\,0918$+$1636 to the boundary of
the colour space where QSOs are selected in SDSS is consistent with
previous studies arguing that metal-rich DLAs most likely are
systematically missing from current DLA samples from optical QSO
surveys (e.g., \citealt{Pei99} and references therein;
\citealt{Noterdaeme09b,Noterdaeme10,Pontzen09}).

In conclusion, our results support the suggestion of F10 that
high-metallicity DLAs are associated with bright galaxy counterparts.
The $z_{\rm abs}=2.5832$ DLA has a high metallicity (for its redshift)
of close to solar despite probing a region about 16 kpc away from the
central emitting region of the DLA galaxy counterpart. We establish
the presence of H$_2$ absorption, and also dust grains based on the
depletion of refractory elements and from reddening of the background
QSO.

\section*{Acknowledgments}

The Dark Cosmology Centre is funded by the DNRF. We thank our
anonymous referee for a fast and constructive report and K. Schmidt
and C. P\'eroux for useful discussions. PN is supported by a
CONICYT/CNRS fellowship. PL acknowledges funding from the Villum
Foundation. JRM is a Sophie \& Tycho Brahe fellow. ST is supported by
the Lundbeck foundation.

\def\aj{AJ}
\def\araa{ARA\&A}
\def\apj{ApJ}
\def\apjl{ApJ}
\def\apjs{ApJS}
\def\apss{Ap\&SS}
\def\aap{A\&A}
\def\aapr{A\&A~Rev.}
\def\aaps{A\&AS}
\def\mnras{MNRAS}
\def\nat{Nature}
\def\pasp{PASP}
\def\aplett{Astrophys.~Lett.}

\bibliographystyle{mn}
\bibliography{thebib}

\end{document}